\documentclass[review, 3p, times]{elsarticle}

\usepackage{booktabs}
\usepackage{amssymb}
\usepackage{amsmath}
\usepackage{bm}
\usepackage{xcolor}
\usepackage[bookmarks=false,colorlinks]{hyperref}
\usepackage{float}
\usepackage{microtype}
\usepackage{lmodern}

\newcommand{\ar}[1]{\textcolor{black}{#1}}
\newcommand{\mjw}[1]{\textcolor{black}{#1}}
\newcommand{\jmr}[1]{\textcolor{black}{#1}}
\newcommand{\rev}[1]{\textcolor{black}{#1}}

\journal{Scripta Materialia}

\newcommand{\kcap}{\mathbf{\hat{k}}}
\newcommand{\comp}{c_{\mathrm{Rh}}}
\newcommand{\cmean}{\langle c \rangle}

\begin{document}

\begin{frontmatter}



\title{Percolation Diagrams Derived from First-Principles Investigation of Chemical Short-Range Order in Binary Alloys}


\author[NU]{Abhinav Roy} 
\author[ASU]{Karl Sieradzki}
\author[NU]{Michael J.\ Waters}
\author[NU]{James M.\ Rondinelli}
\author[NU]{Ian McCue}

\affiliation[NU]{organization={Department of Materials Science and Engineering, Northwestern University},
            addressline={}, 
            city={Evanston},
            postcode={60208}, 
            state={Illinois},
            country={USA}}
\affiliation[ASU]{organization={Ira A.\ Fulton School of Engineering, Arizona State University},
            addressline={}, 
            city={Tempe},
            postcode={85287}, 
            state={Arizona},
            country={USA}}
\begin{abstract}
Recent developments in the percolation theory of passivation have shown that chemical short-range order (SRO) affects the aqueous passivation behavior of alloys. However, there has been no systematic exploration to quantify these SRO effects on percolation in real alloys. In this study, we quantify the effects of SRO on percolation in a binary size-mismatched Cu-Rh alloy and study the related passivation behavior. We develop a mixed-space cluster expansion model trained on the mixing energy calculated using density functional theory. We use the cluster expansion model to sample the configuration space via variance-constrained semi-grand canonical Monte Carlo simulations and develop SRO diagrams over a range of compositions and temperatures. Building on this with the percolation crossover model, specifically the variation of percolation threshold with SRO in the FCC lattice, we construct the first nearest-neighbor chemical percolation diagram. This diagram can inform the design of the next generation of corrosion-resistant metallic alloys.
\end{abstract}


\end{frontmatter}


Chemical short-range order (SRO) in complex concentrated alloys has been demonstrated to affect various properties \cite{taheri2023understanding, chen2022short}, including mechanical \cite{zhang2020short, zhu2023effects, pei2025rigorous} and corrosion behavior \cite{blades2024tuning}. The impact of SRO in corrosion processes has been particularly apparent during primary passivation, where the critical composition of the passivating component can be explained by applying percolation theory \cite{xie2021percolation}. Recent work by the authors has demonstrated that short-range clustering in face-centered cubic (FCC) lattice lowers the percolation threshold of the passivating component in binary alloys, thereby reducing the critical concentration for passivation and improving the corrosion resistance \cite{roy2024effect}. However, there has been no systematic investigation of the effects of SRO on percolation behavior in alloys. Complicating matters further, only a handful of experimental systems have been explored in binary alloys owing to the constraint of needing a wide composition range over which the passivating component exists in solid solution with the base element \cite{xie2021percolation}.

A recent study was conducted on the Cu-Rh binary alloy system \cite{xie2021passivation}, which possesses a solid solution at elevated temperatures, enabling systematic exploration of the entire Cu$_{\mathrm{x}}$Rh$_{\mathrm{1-x}}$ composition space. It was demonstrated that Cu-Rh alloys follow the selective dissolution-based percolation theory of passivation \cite{xie2021passivation, xie2021percolation, roy2024effect}, where the critical Rh content can be linked to the spatial percolation of a Rh-oxide network. Based on the fit of the percolation-crossover model to the experimental data, it was hypothesized that the presence of SRO influences the passivation behavior in the Cu-Rh system. This previous study motivated the investigation in this study to assess the percolation and the associated passivation behavior in these alloys.

Warren-Cowley SRO parameters have been used to describe chemical SRO in alloys, which can be calculated from either experiments \cite{cowley1950approximate, cowley1960short, fantin2020short, taheri2023understanding, wu2025observation} or first-principles simulations \cite{Wolverton1995, Wolverton1998, wolverton2000short}. The Warren-Cowley SRO parameters for a binary system, which are a function of composition $c_B$ and temperature $T$, can be written as:
\begin{equation}
    \alpha_{m} = 1 - \frac{P(B|A)}{c_{B}} 
    \label{eq1}
\end{equation}
where, $P(B|A)$ is the conditional probability of finding a $B$ atom around an $A$ atom in the $m^{\mathrm{th}}$ coordination shell and $c_{B}$ denotes the composition of the $B$ constituent. 
Computational approaches 
to study SRO include 
either performing hybrid molecular dynamics and Monte Carlo simulations using machine learning interatomic potentials \cite{cao2024capturing,smith2024competition}, or using the cluster expansion (CE) method to sample the configuration space and calculate the ensemble average thermodynamic observable of interest, e.g., $\alpha_{m}$  
\cite{sanchez1984generalized,ekborg2024construction,kadkhodaei2021cluster,aangqvist2019icet, van2009multicomponent}. 
The CE method 
maps the configuration-dependent energy of a crystalline system onto a generalized Ising Hamiltonian. The CE model is trained by calculating the observable energy, in this case, the mixing energy of the alloy for a set of symmetrically distinct reference structures \cite{hart2008algorithm,hart2009generating} using density functional theory (DFT). 
\ar{The CE method has been used successfully to determine SRO and the phase diagram of various alloy systems, such as transition metal binary alloys \cite{lu1994first, chinnappan2016first, smith2024competition}, semiconductor binary alloys \cite{wolverton1995first}, and ternary systems \cite{rahm2021tale} by coupling the CE model with Monte Carlo simulations via the canonical or the Variance-Constrained Semi-Grand Canonical (VCSGC) ensembles \cite{sadigh2012calculation}}. 

The real-space CE formalism, however, is inadequate for considering the long-range strain effects that arise due to lattice size mismatch between constituent elements in a coherent single-phase solid solution. For modeling long-range strain effects, the \emph{mixed-space}-CE (MSCE) formalism has been proposed. This formalism incorporates the coherency strain energy, defined as the elastic energy per atom due to lattice mismatch across a coherent interface, into the CE Hamiltonian.
The MSCE formalism has been applied to size-mismatched substitutional systems \cite{Laks1992}, such as Cu-Au, Cu-Ag, and Ni-Au \cite{wolverton1998first}. \ar{The coherency strain energy previously used to analyze the phase stability of epitaxial films and superlattices \cite{ozolins1998effects} also provides insights into the energy penalty for interface formation between unlike constituents due to short-range clustering in the single-phase solid solution.} This leads to the opposite trend in short-range and long-range ordering in certain alloys such as the Cr-Mo-W system \cite{smith2024competition}. In such systems, the SRO tendency is observed in the disordered single-phase solid solution region of the phase diagram, although the system phase separates in the long range (at lower temperatures below the order-disorder transformation temperature). \ar{For the Cu-Rh system, the equilibrium atomic volume of Cu is $7.11$ $\mathrm{cm^{3}mol^{-1}}$ and that of Rh is $8.28$ $\mathrm{cm^{3}mol^{-1}}$ \cite{singman1984atomic}, which results in a volume difference of about $16\%$ due to a significant size mismatch.} Therefore, to achieve the highest accuracy, it is necessary to use the MSCE formalism in the Cu-Rh system, incorporating both the energetics of short-range interactions and long-range coherency strain effects.  

In this study, we performed Monte Carlo simulations in the VCSGC ensemble over a range of temperatures and compositions, thereby developing chemical SRO diagrams. The computed SRO parameters can be used to determine the percolation threshold for the system with a given degree of SRO by using a polynomial fit that we have derived previously \cite{roy2024effect}. We can therefore assess the deviation of the percolation threshold from that of a random solid solution. From the SRO diagrams, using the polynomial fit that relates the percolation threshold in the FCC lattice to SRO parameters, we establish a subsequent \emph{chemical percolation diagram}. Using this diagram, we can infer the passivation behavior of Cu-Rh alloys and also demonstrate how SRO can be used as a `knob' to tune the corrosion resistance. 

The training structures for the CE model were relaxed using DFT, as implemented in the Vienna Ab initio Simulation Package (VASP) \cite{kresse1996}. Parameters for the structure-relaxation calculations were chosen from relevant convergence tests (details can be found in Section S1 of the Supplementary Materials (SM)). \ar{The calculations were set up and coupled with the CE implementation using the Atomic Simulation Environment \cite{Larsen2017}.} In this study, we utilized the ICET \cite{aangqvist2019icet} Python package to calculate coherency strain energy and implement the 
MSCE model \cite{wolverton2000short, wolverton1998first, Laks1992, ozolins1998effects, Blum2004, rahm2022quantitative}.  \ar{The effective Hamiltonian for the MSCE 
can be written as follows:}
\begin{equation}
\mathcal{H} = J_{0} + m_{\beta}J_{\beta} \left< \Gamma_{\beta'}(\bm \sigma) \right> _{\beta} +\ \mathcal{E}_{CS}
\label{eq2}
\end{equation}
\autoref{eq2} consists of point and multibody interactions in the real space, where $\beta$ represents an orbit (set of symmetrically equivalent clusters). \ar{The term $J_{0}$ in the equation is a constant also called a zerolet in the CE literature.} Here, $\bm \sigma$ is a vector describing the configuration of the constituents over all the lattice points. The orbit $\beta$ can be of different orders, such as 2nd order (pairs), 3rd order (triplets), and so on. In \autoref{eq2}, $m_{\beta}$ represents the multiplicity of each orbit, $J_{\beta}$ represents the effective cluster interaction parameters (contribution of clusters to the model), and $\langle\Gamma_{\beta'}(\bm \sigma)\rangle_{\beta}$ represents the average over respective cluster functions belonging to the orbit $\beta$. The term $\mathcal{E}_{CS}$ represents the strain energy that consists of the coherency strain energy $\Delta E_{CS}(\mathbf{\hat{k}},c)$, which is a function of composition and the crystallographic direction $\kcap$ representing the interface normal between the constituents. The strain energy is evaluated in reciprocal space because the long-range strain fields decay much more slowly with distance than multi-body interactions associated with clusters. Details of the strain energy calculation, hyperparameter optimization, and DFT-CE mixing energy comparison are given in the SM.

We used the MSCE model for Cu-Rh alloy to sample the configurational space via Monte Carlo (MC) simulations in the VCSGC ensemble.
A supercell containing 4000 atoms was used for the MC simulations, and we performed calculations for 400 Monte Carlo Sweeps (MCS), where $1$ MCS = $4000$ MC steps (equivalent to $1$ MC step per atom). We discard the first 200 \mjw{(first half)} of the MCS as an equilibration run. Thereafter, we calculated the Warren-Cowley SRO parameters by taking the average over the remaining 200 \mjw{(last half)} of the MCS, which was considered as the production run. 
We note that the correlation length depends on composition and temperature, but 1 MCS was found to be a sufficient sampling distance for the entire composition range (see Section S5 of SM for more details).

We aim to obtain SRO diagrams in the single-phase solid solution region of the Cu-Rh binary phase diagram; however, the phase boundary predicted by the MSCE model may not align with the experimental bulk phase diagram, owing in part to missing vibrational entropy contributions. 
%
The MC simulation does,
 however, capture configurational contributions to entropy, which we used to ensure the SRO diagrams were captured in the single-phase solid solution regime. 
%
One advantage of performing MC simulations in the VCSGC ensemble is that the Helmholtz free energy can be derived from the parameters of the ensemble given by the following expression:
\begin{equation}
    \frac{\partial F}{\partial c} = -Nk_{B}T\kappa(\phi + 2\cmean)\,,
\end{equation}
\ar{where the dimensionless parameters $\phi$ constraints the mean of concentration ($\cmean$) and $\kappa$ constraints the variance. The total number of sites $N$ is held constant in the ensemble, and for large enough $\kappa$, the probability of a microstate peaks at $\cmean = -\phi /2$.} This allows for thermodynamic integration of free energy, thereby allowing for mapping the phase boundary.

\begin{figure*}
	\centering
	\includegraphics[width=\textwidth]{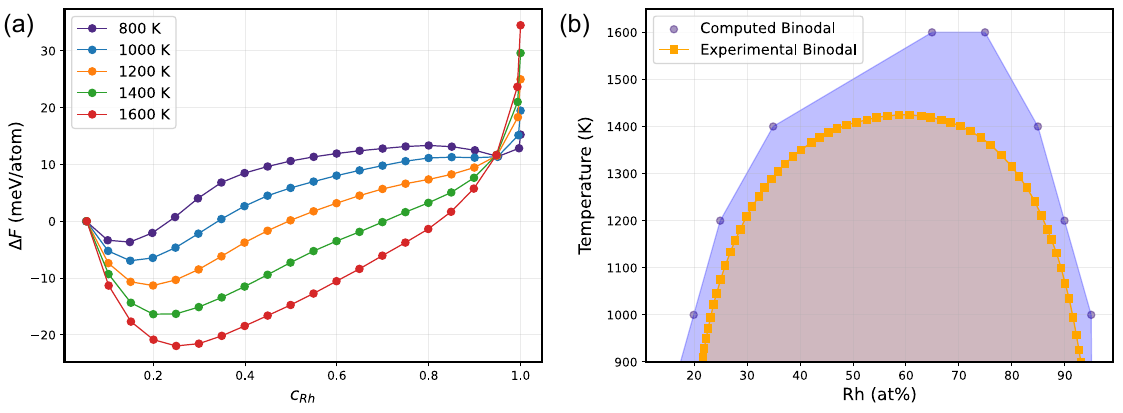}\vspace{-20pt}
	\caption{(a) Helmholtz free energy obtained via thermodynamic integration. (b) Comparison of the computed and experimental \cite{chakrabarti1982cu} Cu-Rh binary phase diagram.}\label{binodal}
\end{figure*}

We performed MC simulations across a range of temperatures ($T$: 800-1600\,K) and the entire composition range. 
The experimental Cu-Rh phase diagram exhibits a miscibility gap, and a precursor to this behavior can be seen in the free energy curves for different temperatures, \autoref{binodal}(a), such as the double-well shape observed in the curve at 800\,K.
The solubility limits and miscibility gap are readily extracted via the common tangent construction, \autoref{binodal}(b), and we overlaid the experimental binodal obtained from Ref.\  \cite{chakrabarti1982cu} for comparison. 
\rev{We find that the computed order-disorder transformation temperature from our calculations is 1600 K and that obtained from experiments is 1423 K. The 177 K temperature difference is attributed to the absence of vibrational entropy contributions to the free energy in our phase diagram calculation, which is a limitation of the MSCE model. Therefore, it is important to note that the phase boundaries and the SRO diagrams are derived from a configuration-only free energy model. Various methods have been proposed in the literature, such as the transferable force-constants method \cite{van2002effect} or, more recently, the use of Machine Learning Force Fields to calculate the vibrational free energy \cite{tolborg2023low}. However, it was pointed out that in the case of the Cr-Mo-W system \cite{smith2024competition}, the magnitude of SRO parameters calculated using real-space CE (without vibrational effects) and the moment tensor potential (including vibrational effects) are comparable. Moreover, such effects are more important in systems with dynamical instabilities. Including this effect is essential for high-fidelity phase diagram simulations and may affect the magnitude of short-range order near the phase boundaries; however, this analysis is outside the scope of the paper.}

\begin{figure*}
\centering
\includegraphics[width=0.8\textwidth]{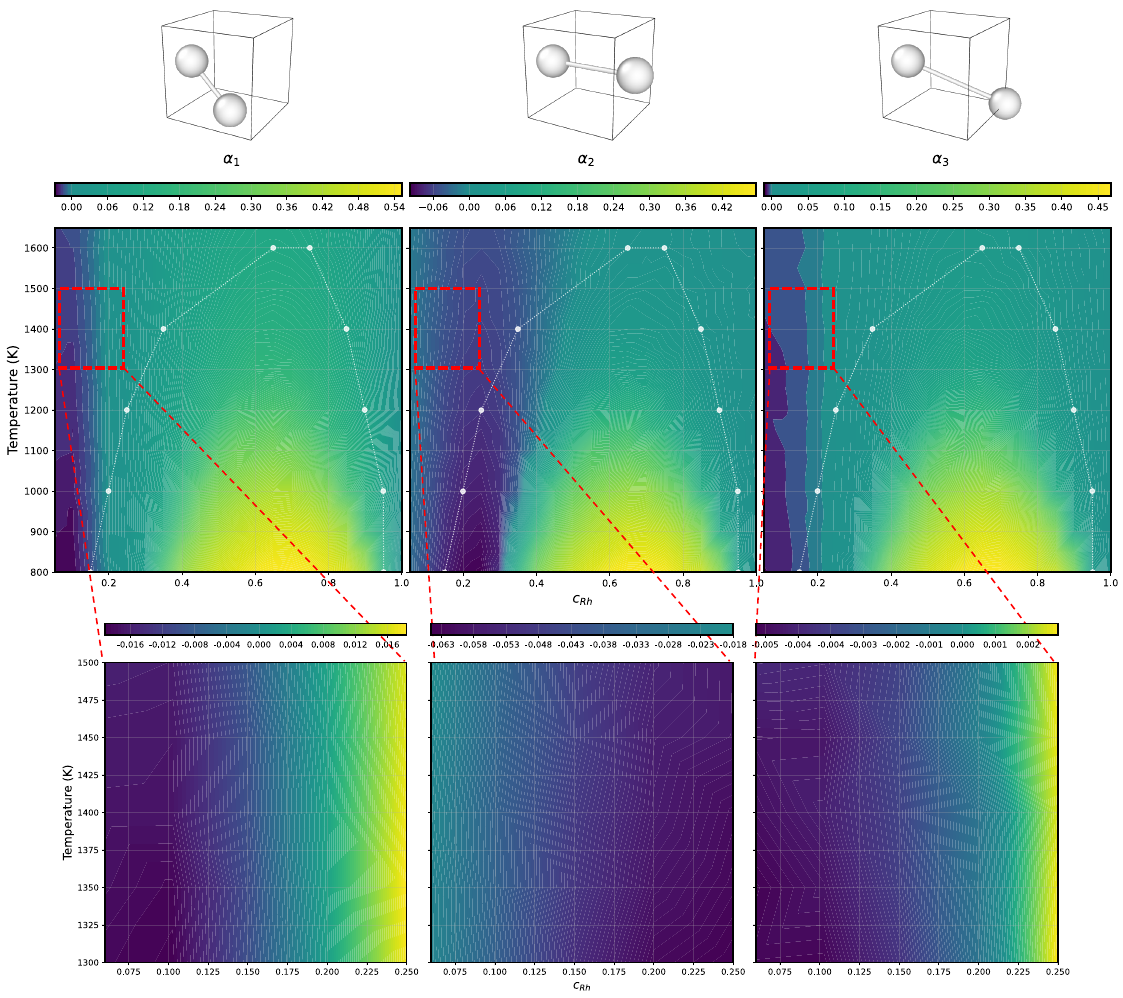}\vspace{-10pt}
\caption{SRO diagrams for (left) first, (center) second, and (right) third nearest-neighbor Warren-Cowley $\alpha$ parameters. The computed binodal is overlaid on the SRO diagrams to delineate the single-phase solid solution region (highlighted by the red box).}\label{sro_map}
\end{figure*}

From the computed phase diagram, we demarcate the single-phase solid solution region of the MSCE model corresponding to the composition range of $0.05 \le \comp \le 0.25$, where $\comp = \cmean$ and temperature range $1300 \le T \le 1500$ K. \rev{The experimental solidus temperature of Cu-Rh solid solution is in the range of 1433 K to 1543 K for Rh composition between 7-25 at\% \cite{chakrabarti1982cu}. The temperature range chosen for the single-phase solid solution falls within the Monte Carlo simulation range, and it is below the maximum experimental solidus temperature within the composition of interest. Therefore, the temperature chosen for our computational analysis is experimentally relevant.} It was observed that the SRO parameters are microstate dependent. We thereby obtained a distribution of SRO parameter fluctuations about the ensemble average (Figure S12 of SM). We used this distribution to calculate the standard error of the mean. We found that sampling $200$ data points to report the ensemble average of $\alpha$ values is adequate, as it keeps the relative standard error within $10\%$ for the first nearest neighbor (NN) SRO parameter.

Next, we calculated the Cu-Rh SRO diagrams for the first three NN Warren-Cowley SRO parameters (\autoref{sro_map}). From the diagrams, we see that the signs of the SRO parameters alternate from the first NN to the second NN, and then to the third NN. This modulation is consistent with the trend of SRO parameters proposed experimentally \cite{xie2021passivation}, but this alternating trend is observed above the theoretical first NN percolation threshold of the FCC lattice ($p_{c}^{3D}\{1\} = 0.1992$ \cite{lorenz2000similarity}).

Despite clustering being the expected behavior for systems with miscibility gaps, the Cu-rich phase shows an ordering tendency. This result is attributed to the energy penalty of interface formation due to coherency strain energy, and explained in terms of coherent spinodal depression in systems with a miscibility gap \cite{cahn1962coherent}. To the best of our knowledge, this mismatch in short-range and long-range ordering trend has not been reported for the Cu-Rh system before. Recently, it has been demonstrated that short-range ordering in alloys can also be influenced by magnetism. The spin interaction between ferromagnetic Fe and Ni, and anti-ferromagnetic Cr, leads to a significant deviation in the SRO values obtained from MC simulation using a CE model \cite{su2024first}. However, in this study, we have considered diamagnetic Cu and weakly paramagnetic Rh. Therefore, we do not consider the effect of magnetism on SRO, and such an analysis is beyond the scope of this study.

\ar{The first NN percolation threshold values (denoted by $p_{c}^{3D} \{1\}$)} for different values of first NN Warren-Cowley $\alpha_{1}$ parameters were obtained using the Large Cell Monte Carlo Renormalization Group (MC-RNG) method (details of which can be found in Ref.\  \cite{roy2024effect}). 
\ar{We fit a third degree polynomial to the data (Figure S14 of SM) given by $p_{c}^{3D}\{1\} = p_{c,0}^{3D} + \nu_{1}\alpha_1 + \nu_{2}\alpha_{1}^{2} + \nu_{3}\alpha_{1}^{3}$, 
where $p_{c,0}^{3D}$ represents the percolation threshold for a random alloy ($\alpha=0$). The polynomial fit can be used to interpolate the percolation threshold values for the given range of $\alpha_1$ values without performing the computationally expensive MC-RNG calculations. The confidence-of-fit is $R^2=0.998$ for the polynomial fit with $\nu_{1} = -0.0582$, $\nu_{2} = -0.0396$, $\nu_{3} = 0.0534$, respectively.}
\jmr{It is important to note that the percolation threshold in a lattice of fixed dimensionality is a unique value. However, when percolation is applied to alloys, i.e., where the percolating units are the chemical constituents, the threshold can vary due to SRO.} \ar{The preferential local arrangement of atoms due to the enthalpic and entropic effects leads to deviation from the Bernoulli percolation threshold of the random lattice (where the site occupation probability is truly random).} 
\begin{figure*}[ht]
\centering
\includegraphics[width=\textwidth]{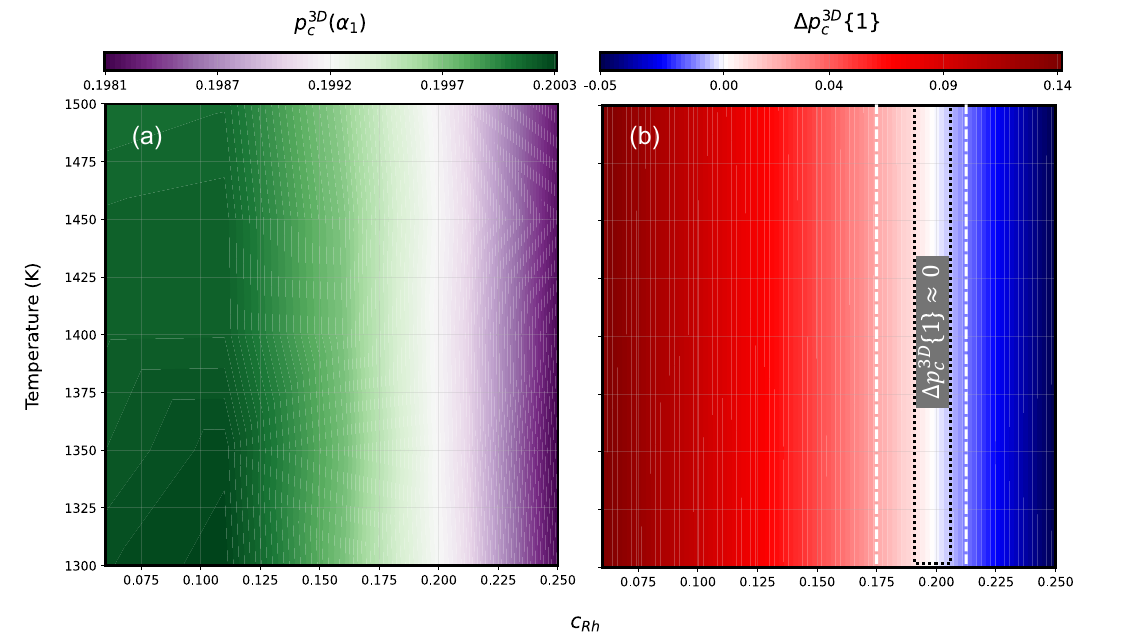}\vspace{-20pt}
\caption{Chemical percolation diagram (CPD) for the Cu-Rh alloy. (a) The percolation threshold diagram of CPD obtained from the polynomial fit. (b) The percolation threshold difference diagram. The region within the composition range of $0.175\leq\comp\leq0.21$, represented by the dashed lines, delineates the range of maximum possible variation in $ p_{c}^{3D}(\alpha_1)$ based on the values of $\alpha_1$ used for the MC-RNG analysis (Figure S14 of SM).} \label{percolation_map}
\end{figure*}

In addition to the first NN percolation threshold for Cu-Rh shown in \autoref{percolation_map}(a), we plot the difference between calculated percolation threshold and alloy composition in \autoref{percolation_map}(b), denoted by $\Delta p_{c}^{3D} \{1\} = p_{c}^{3D}(\alpha_1) - \comp$, where, $p_{c}^{3D}(\alpha_1) \coloneqq p_{c}^{3D}\{1\}$. We note that in the experimental study of Cu-Rh alloy \cite{xie2021passivation}, the data for the number of monolayers selectively dissolved ($h$) for passive film formation plotted against the alloy composition ($\comp$) was fit to the percolation-crossover model corresponding to the third NN percolation threshold $p_{c}^{3D} \{1,2,3\}$ (Figure 7 of Ref.\  \cite{xie2021passivation}). This is due to the atomic size of Cu and Rh atoms that leads to an estimated \{-Rh-O-Rh-\} oxide mer unit length close to the third NN distance in the FCC lattice. \rev{Therefore, the third NN percolation threshold is relevant for determining the critical concentration for passivation for binary Cu-Rh. In principle, it is possible to extend the MC-RNG analysis to the third NN percolation threshold, though it would be computationally expensive. From \autoref{percolation_map}(a), we can infer that the percolation threshold increases due to short-range ordering. Based on the third NN short-range order diagram (\autoref{sro_map}), we can expect that the third NN percolation threshold will show a similar trend. This behavior in percolation threshold shifting to a higher value is also observed in the percolation crossover model fit to the experimental data in Ref.\  \cite{xie2021passivation}.}

\ar{The chemical species act as the percolating units; accordingly, the diagram shown in \autoref{percolation_map} is referred to as a chemical percolation diagram (CPD).
%
The CPD comprises two distinct diagrams: the first is the percolation threshold diagram, shown in \autoref{percolation_map}(a), and the second is the percolation threshold difference diagram, shown in \autoref{percolation_map}(b). Together, these diagrams form the chemical percolation diagram (CPD) for the Cu--Rh alloy.} 
Similarly, CPD can be constructed from first-principles for other alloy systems.
%
The percolation threshold difference diagram of the CPD shows a region of temperature and composition where the calculated percolation threshold due to SRO is approximately equal to the composition of the percolating constituent.

We name this region in the percolation threshold difference diagram as the \textit{critical} region. Any region to the left of the \textit{critical} region represents alloys that will never have a spanning (percolating) network of the constituent, irrespective of the composition and temperature chosen in this region. Accordingly, any region towards the right will always contain a percolating network. The \textit{critical} region determines the composition at which percolation network formation begins during selective dissolution in alloys with SRO. Shifting this region to the left, by employing appropriate alloy processing protocols, can enhance the passivation behavior of the alloy. 

\rev{We note that the CPD developed in this study for Cu-Rh binary alloy corresponds to the first NN percolation threshold. Therefore, the CPD in \autoref{percolation_map} cannot be directly applied to the simple binary Cu-Rh system. It would be necessary to extend the analysis to third NN percolation and derive the corresponding CPD. The methodology for creating CPD for alloys is the main contribution of this study; however, the CPD developed here can serve as a design tool for complex concentrated alloys (CCA) with Rh as the passivating component. Moreover, from a materials design perspective, the CPD corresponding to the 1st NN is most important, as SRO can be used as a tuning parameter to alter the percolation threshold significantly. The NN spacing can be tuned by varying the composition and the number of non-passivating components except Rh. Therefore, designing CCA by tuning the number of non-passivating components and their respective composition in such a manner that Rh-O-Rh mer unit spacing becomes comparable to the 1st NN spacing of the FCC lattice will make the analysis in the paper directly applicable to such alloys \cite{cantor2004microstructural}.}

To conclude, we developed a mixed-space cluster expansion model for the size-mismatched Cu-Rh binary alloy to investigate the percolation and the associated passivation behavior. The coherency strain energy was explicitly included in the cluster expansion model. Using the cluster expansion model to calculate the Warren-Cowley SRO parameters via Monte Carlo simulations, we demonstrated that the Cu-Rh alloy exhibits an opposite short-range ordering tendency compared to its long-range phase-separating behavior. The polynomial fit of the percolation threshold variation with $\alpha_1$ from our percolation-crossover model was used to obtain the associated first NN chemical percolation diagram for the Cu-Rh alloy. We identified a \textit{critical} region in the chemical percolation diagram that sets the composition for the onset of percolation in binary alloys. From a design perspective for complex concentrated alloys, this diagram can serve as a tool to guide element selection and alloy processing, allowing the \textit{critical} region to be shifted towards the left and facilitating the design of corrosion-resistant materials.

\section*{CRediT authorship contribution statement}
\textbf{Abhinav Roy:} Conceptualization, Data curation, Formal Analysis, Investigation, Methodology, Software, Visualization, Writing - original draft.
\textbf{Karl Sieradzki:} Writing – review \& editing,
Supervision, Funding acquisition, Conceptualization.
\textbf{Michael J. Waters:} Methodology, Software, Writing – review \& editing.
\textbf{James M. Rondinelli:} Writing – review \& editing,
Supervision, Funding acquisition, Conceptualization.
\textbf{Ian McCue:} Writing – review \& editing,
Supervision, Funding acquisition, Conceptualization.
\section*{Declaration of competing interest}
The authors declare that they have no known competing financial
interests or personal relationships that could have appeared to influence
the work reported in this paper.
\section*{Acknowledgments}
This work was supported by the National Science Foundation (NSF) under award numbers DMR-2208865 (A.R., J.M.R., and I.D.M.) and DMR-2208848 (K.S.). A.R. gratefully acknowledges support from the Ryan Fellowship and the International
Institute for Nanotechnology at Northwestern University. 
M.J.W.\ acknowledges funding from the Office of Naval Research (ONR) through the Multidisciplinary University Research Initiative (MURI) program (award number N00014-20-1-2368).
This research was supported in part through the computational resources and staff contributions provided for the Quest high performance computing facility at Northwestern University which is jointly supported by the Office of the Provost, the Office for Research, and Northwestern University Information Technology. This work used Bridges-2 at Pittsburgh Supercomputing Center through allocation mat230009p from the Advanced Cyberinfrastructure Coordination Ecosystem: Services \& Support (ACCESS) program, which is supported by National Science Foundation grants \#2138259, \#2138286, \#2138307, \#2137603, and \#2138296.
\section*{Supplementary materials}
Supplementary material associated with this article can be found, in
the online version, at \url{http://dx.doi.org/10.1016/j.scriptamat.2025.117137}
\bibliographystyle{elsarticle-num} 
\bibliography{refs_paper2.bib}

\end{document}


\begin{singlespace}
\title{\textsc{supplementary  materials}\\[1em] Percolation Diagrams Derived from First-Principles Investigation of Chemical Short-Range Order in Binary Alloys}
\author[a]{Abhinav Roy}
\author[b]{Karl Sieradzki}
\author[a]{Michael Waters}
\author[a]{James Rondinelli}
\author[a]{Ian McCue}

\affil[a]{Department of Materials Science and Engineering, Northwestern University, Evanston, 60208, Illinois, USA}
\affil[b]{Ira A.\ Fulton School of Engineering, Arizona State University, Tempe, 85287, Arizona, USA}
\date{}
\maketitle
\tableofcontents
\end{singlespace}
\clearpage
\section{Details of density functional theory calculations}
\begin{figure}[H]
    \centering
    \includegraphics[width=\textwidth]{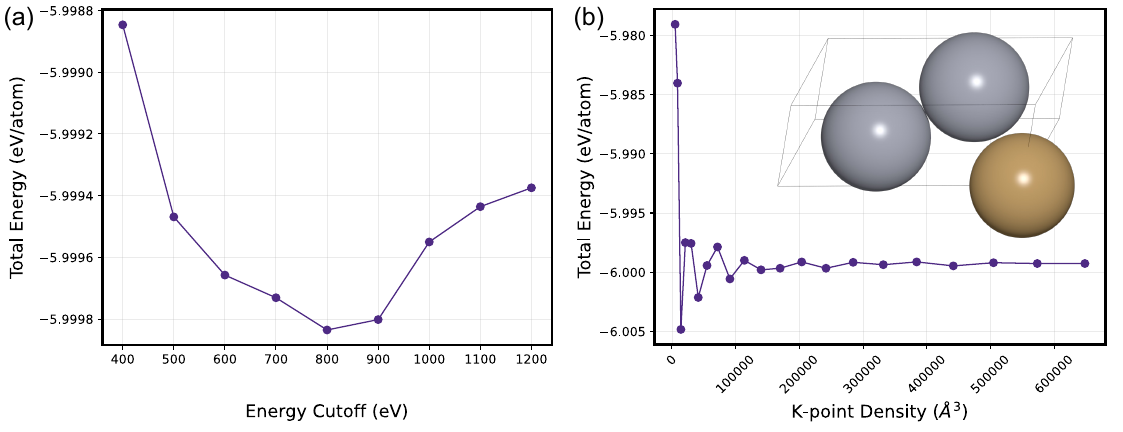}\vspace{-10pt}
    \caption{Convergence tests for determining (a) the optimal plane-wave energy cutoff and (b) k-point density (without the 1/8$\pi^3$ multiplication factor).} \label{conv_test}
\end{figure}

\ar{We used the projector augmented wave method \cite{kresse1999} with an optimal value of plane-wave energy cutoff and the generalized gradient approximation, as implemented by Perdew, Burke, and Ernzerhof for the exchange-correlation functional \cite{perdew1996}.} For accurate calculations, we used an energy convergence of $10^{-9}$ eV and geometry optimization was converged with force tolerance of $10^{-3}$ eV\,\AA$^{-1}$  per atom. The Methfessel-Paxton smearing scheme of order one was used with a smearing width of $0.05$ eV. We conducted convergence tests for the plane-wave energy cutoff and the k-point density to determine the optimal values of these parameters for the structure relaxations. First, we performed convergence tests for several structures with varying amounts of Cu and Rh atoms. \autoref{conv_test}(a) shows the energy cutoff convergence test for a structure with one Cu atom and two Rh atoms in the primitive cell (with k-point density more than 25000/$\pi^3$ \AA$^3$). We chose the optimal k-point density and plane-wave energy cutoff based on the criterion that the variation in total energy is less than $4$ meV/atom, as this becomes the bottleneck for the accuracy of our cluster expansion model. Based on this criterion, we obtained a plane-wave energy cutoff of 600 eV. Thereafter, we determined the optimum k-point density for the same structures with an energy cutoff of 600 eV. For the previously mentioned structure with one Cu and two Rh atoms (shown in the inset of \autoref{conv_test}(b)), the convergence test for k-point density is shown in \autoref{conv_test}(b). We note that the x-axis label in \autoref{conv_test}(b) must be multiplied by a factor of 1/8$\pi^3$ to get the actual k-point density. Based on the total energy convergence criterion,  we sampled the Brillouin zone for structures with a $\Gamma$-centered $k$-point mesh corresponding to $k$-point density of $25000/\pi^{3}$ \AA$^{3}$.
\pagebreak
\section{Mathematical formulation of epitaxial strain energy along \texorpdfstring{$\kcap$}{k}}
\begin{figure}[h]
    \centering
    \includegraphics[width=\textwidth]{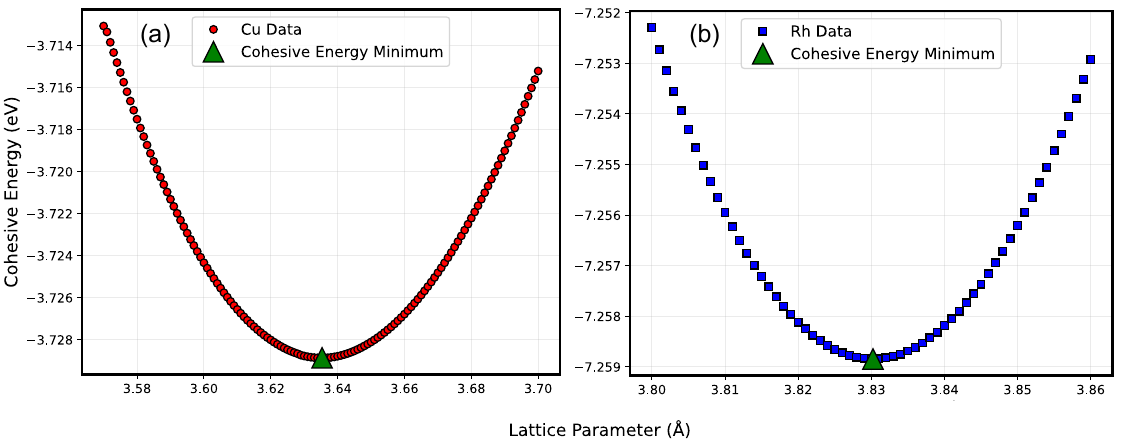}\vspace{-10pt}
    \caption{Cohesive energy of (a) Cu and (b) Rh primitive cell plotted for various lattice parameters. The cohesive energy minima labeled in the figures correspond to the equilibrium lattice parameter calculated using DFT.} \label{cohesive_energy}
\end{figure}

We describe the mathematical formulation for the calculation of epitaxial strain energy as implemented in the ICET Python package \cite{aangqvist2019icet, rahm2022quantitative, ozolins1998effects}. We first calculate the equilibrium lattice parameter (corresponding to the minimum cohesive energy) from a standard equation of state calculation using density functional theory (DFT), as shown in \autoref{cohesive_energy}.
The equilibrium lattice parameters for Cu ($a_{\mathrm{Cu}}$) and Rh ($a_{\mathrm{Rh}}$) is found to be $3.6353$ \AA\ and $3.8302$ \AA, respectively. We define the interface normal vector $\kcap$ such that the system relaxes along this direction while deforming biaxially along the perpendicular directions $\mathbf{k}_{\perp1}$ and $\mathbf{k}_{\perp2}$. To get the epitaxial energy, we perform a lattice scaling transformation along the direction of the unit vector $\kcap$. The unit vector can be obtained from the normalization of $\mathbf{k} = (k_x, k_y, k_z)$ as follows:
\begin{equation}
\kcap = \frac{\mathbf{k}}{||\mathbf{k}||}
\end{equation}
We can define the two orthogonal vectors $\mathbf{k}_{\perp1} = (k_{\perp1,x}, k_{\perp1,y}, k_{\perp1,z})$ and $\mathbf{k}_{\perp2} = (k_{\perp2,x}, k_{\perp2,y}, k_{\perp2,z})$ by projecting an arbitrary vector $\mathbf{v} = (v_x,v_y,0)$ onto the plane perpendicular to $\kcap$ such that $k_{\perp1,x} = v_x$ and $k_{\perp1,y} = v_y$. By enforcing orthogonality we can write $k_{\perp1,z}$ as follows:
\begin{equation}
k_{\perp 1, z} = - \frac{k_{\perp 1, x} k_x + k_{\perp 1, y} k_y}{k_z}
\end{equation}
We normalize $\mathbf{k}_{\perp 1}$ to $\mathbf{\hat k}_{\perp 1}$ and obtain another perpendicular orthonormal vector as follows:
\begin{equation}
    \mathbf{\hat k}_{\perp 2} = \mathbf{\hat{k}} \times \mathbf{\hat k}_{\perp 1}
\end{equation}
These vectors can be combined to obtain an orthonormal basis $X$ as follows:
\begin{equation}
X = [\mathbf{\hat{k}},\; \mathbf{\hat k}_{\perp 1},\; \mathbf{\hat k}_{\perp 2}]
\end{equation}
Thereafter, we can define a scaled basis $X'$ using scale factors $s_\parallel$ and $s_\perp$ as follows:
\begin{equation}
X' = [s_{\parallel} \mathbf{\hat{k}},\; s_{\perp} \mathbf{\hat k}_{\perp 1},\; s_{\perp} \mathbf{\hat k}_{\perp 2}]
\end{equation}
Let the FCC lattice vectors be columns in matrix $C$, then the new lattice after scaling is:
\begin{equation}
C' = X' X^{-1} C
\end{equation}
This transformation stretches the lattice parallel to $\kcap$ by $s_\parallel$ and perpendicular to $\kcap$ by $s_\perp$.

To calculate the epitaxial strain energy from first-principles, we perform a series of total energy calculations using DFT to estimate the energies of stretched Cu and Rh lattices. For this, we define a set of perpendicular scale factors for Cu ($s_{\perp,\mathrm{Cu}}$) and Rh ($s_{\perp, \mathrm{Rh}}$) as follows:
\begin{equation}
s_{\perp, \kappa} 
= \left\{
\zeta_{\min,\kappa} + \frac{i}{n - 1} \left(\zeta_{\max,\kappa} - \zeta_{\min,\kappa}\right) 
\ \middle| \ i \in \{0,1,\dots, n-1\}
\right\}, \quad \kappa \in \{\mathrm{Cu},\mathrm{Rh}\}
\label{eq6}
\end{equation}
In \autoref{eq6}, $n$ corresponds to the total number of elements in the set ($n=10$ in our calculations), and the other parameters are given as follows:
\begin{equation}
\zeta_{\min,\mathrm{Cu}} = 0.98, \quad 
\zeta_{\min,\mathrm{Rh}} = \frac{a_{\mathrm{Cu}}}{a_{\mathrm{Rh}}}, \quad
\zeta_{\max,\mathrm{Cu}} = \frac{a_{\mathrm{Rh}}}{a_{\mathrm{Cu}}}, \quad 
\zeta_{\max,\mathrm{Rh}} = 1.02
\end{equation}
For a given $s_{\perp, \kappa}$ in the set defined in \autoref{eq6}, we can define the set of parallel scale factors:
\begin{equation}
s_{\parallel, \kappa} =
\left\{
\beta_{\min, \kappa} + \frac{i}{n-1} \left[ \beta_{\max, \kappa} - \beta_{\min, \kappa}) \right]
\ \middle|\ i \in \{0,1,\dots,n-1\}
\right\}, \quad \kappa \in \{\mathrm{Cu},\mathrm{Rh}\},
\label{eq7}
\end{equation}
Where the bounds are:
\begin{equation}
\beta_{\min, \kappa} = s_\perp^{-0.8} - 0.12,
\quad
\beta_{\max, \kappa} = s_\perp^{-0.8} + 0.12.
\end{equation}

\begin{figure}[H]
	\centering
	\includegraphics[width=\textwidth]{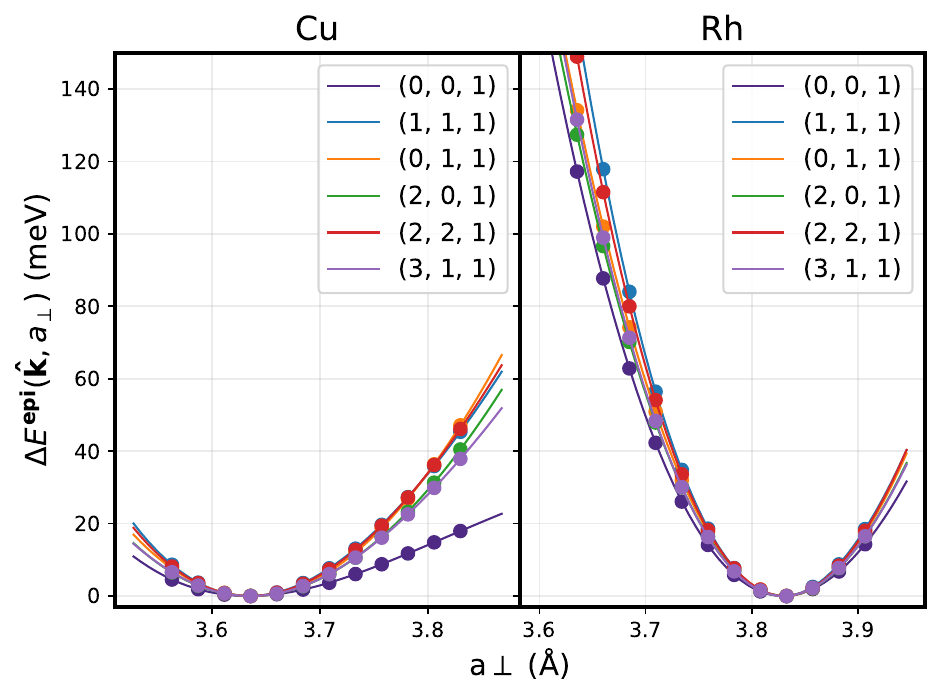}\vspace{-10pt}
	\caption{Epitaxial strain energy for Cu and Rh. The epitaxial energy is calculated as the excess energy with respect to the energy at the equilibrium lattice parameter and subsequently fitted to a quintic polynomial as shown.} \label{epitaxial_strain}
\end{figure}

For each $s_{\perp,\kappa}$ among the set of values in \autoref{eq6} and $s_{\parallel,\kappa}$ in \autoref{eq7}, we sample the energy $E_{\parallel}(a_{\parallel})$ at the following lattice parameters:
\begin{equation}
a_{\parallel,\kappa} = a_\kappa \times s_{\parallel,\kappa}
\end{equation}
Where $a_\kappa$ is the equilibrium lattice parameter for the given element ($\kappa \in \{\mathrm{Cu}, \mathrm{Rh}\}$). Thereafter, we fit a cubic polynomial to the set of energies obtained as follows:
\begin{equation}
E_{\parallel}(a_{\parallel, \kappa}) \approx \sum_{i=0}^{3} p_i a_{\parallel, \kappa}^i
\end{equation}
The minimum energy with respect to $a_{\parallel, \kappa}$ from the cubic polynomial fit is denoted by:
\begin{equation}
E_{\perp}(a_{\perp, \kappa}) = \min\{E_{\parallel}(a_{\parallel, \kappa})\}
\end{equation}
Thereafter, we fit a quintic polynomial to the minimum energy values as follows:
\begin{equation}
E_{\perp}(a_{\perp, \kappa}) \approx \sum_{j=0}^5 c_j a_{\perp, \kappa}^j
\end{equation}
For each direction $\kcap$, the epitaxial strain energy is given by the following expression:
\begin{equation}
\Delta E_{\kappa}^{\mathrm{epi}}(\hat{k}, a_\perp) = E_{\perp}(a_{\perp, \kappa}) - E_{\perp}(a_\kappa)
\end{equation}
The constant $c_{0}$ is adjusted so that epitaxial strain energy is zero at the equilibrium lattice constant, i.e., $\Delta E_{\kappa}^{\mathrm{epi}}(\hat{k}, a_\kappa) = 0$. The epitaxial strain energy for Cu and Rh is shown in \autoref{epitaxial_strain}.
\pagebreak
\section{Details of strain energy calculations}
In this section, we describe the calculation of coherency strain energy $\Delta E_{CS}(\kcap,c)$ for arbitrary crystallographic directions and the strain energy term $\mathcal{E}_{CS}$ that appears in the effective cluster expansion (CE) Hamiltonian. The strain energy term can be written as:
\begin{equation}
\mathcal{E}_{CS} = \frac{1}{4c(1-c)}\sum_{\mathbf{k}}\Delta E_{CS}(\mathbf{\hat{k}},c)|S(\mathbf{\hat{k}}, \bm \sigma)|^{2} e^{(\eta |\mathbf{k}|)^{2}}
\label{eq3}
\end{equation}
In \autoref{eq3}, $S(\kcap,\bm \sigma)$ represents the structure factor and $e^{(\eta |\mathbf{k}|)^{2}}$ is the damping factor to attenuate the strain energy for large $\kcap$ structures.

\begin{figure}[h]
    \centering
    \includegraphics[width=\textwidth]{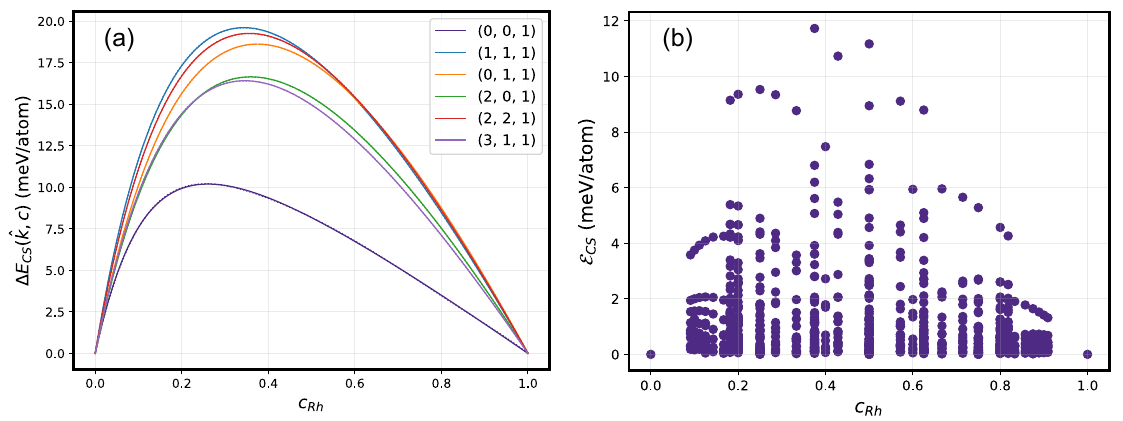} \vspace{-10pt} 
    \caption{\ar{(a) The coherency strain energy for different crystallographic directions over the entire composition range. (b) Strain energy term $\mathcal{E}_{CS}$ calculated for all the training structures of the CE model.}}\label{strain_energy}
\end{figure}

The coherency strain energy $\Delta E_{CS}(\kcap,c)$ can be written as a function of the epitaxial strain energy $\Delta E^{\mathbf{epi}}_{Cu}(\kcap,a_{\perp})$ and $\Delta E^{\mathbf{epi}}_{Rh}(\kcap,a_{\perp})$ as follows:
\begin{equation}
    \Delta E_{CS}(\kcap,c) = \underset{a_{\perp}}{\min}\left[ (1-c)\Delta E^{\mathbf{epi}}_{Cu}(\kcap,a_{\perp}) + (c)\Delta E^{\mathbf{epi}}_{Rh}(\kcap,a_{\perp}) \right]
    \label{eq4}
\end{equation}
The coherent interface will biaxially deform to reach an equilibrium lattice parameter $a_{\perp}$ perpendicular to the interface normal vector $\kcap$. The epitaxial energy $\Delta E^{\mathbf{epi}}_{Cu}(\kcap,a_{\perp})$ and $\Delta E^{\mathbf{epi}}_{Rh}(\kcap,a_{\perp})$ were calculated by performing a series of DFT calculations to obtain the minimum energy parallel to the interface normal for a given $a_{\perp}$.
The $\Delta E_{CS}(\kcap,c)$ (\autoref{eq4}) for six crystallographic directions of the FCC lattice is shown in \autoref{strain_energy}(a), and the strain energy term $\mathcal{E}_{CS}$ (\autoref{eq3}) is shown in \autoref{strain_energy}(b).
The coherency strain energy is a function of both composition and the crystallographic direction corresponding to the interface normal.

To calculate $\Delta E_{CS}(\kcap_{i},c)$ for arbitrary crystallographic directions $\kcap_{i}$, we fit the Redlich-Kister (RK) polynomial \cite{redlich1948algebraic} to the coherency strain energy as: 
\begin{equation}
    \Delta E_{CS}(\kcap_{i},c) = c(1-c)\sum_{n=1}^{M}L^{i}_{n}(\kcap_{i})(1-2c)^{n}\,,
    \label{eq5}
\end{equation}
where the RK coefficients $L^{i}_{n}(\kcap_{i})$ are a function of crystallographic direction. We chose $M=5$ for the degree of RK polynomial. The stereographic projection of the interface normal vector ($\kcap_{i}$) is taken mathematically represented by $\bm{u}_{i} = (x_{i},y_{i})$. A second degree ($d=2$) polynomial basis vector $\lambda (\bm{u}_{i}) \in \mathbb{R}^{B}$, where $B = (d+1)(d+2)/2$, is constructed such that $\lambda (\bm{u}_{i}) = [x_{i}^{p}y_{i}^{q}] \enspace \forall \enspace p,q \in \mathbb{W}$ such that $p+q \leq 2$. The orientation-dependent RK parameters are obtained by solving the optimization problem for the coefficient matrix $\Theta \in \mathbb{R}^{B \times M}$ as 
$    \underset{\Theta}{\mathrm{min}} ||\Lambda \Theta - \mathcal{L}||^{2}_{F}$, 
where $||\cdot||_{F}$ is the Frobenius norm. Here, the RK parameter for a given $\kcap_{i}$ is $L_{n}^{i}(\kcap_{i}) \in \mathbb{R}^{N}$. For $M$ such RK parameters, the basis matrix $\Lambda \in \mathbb{R}^{N \times B}$ and the parameter matrix $\mathcal{L} \in \mathbb{R}^{N \times M}$ is given as follows:
\begin{equation}
\Lambda = \begin{bmatrix}
\lambda(\bm{u}_1) \\
\lambda(\bm{u}_2) \\
\vdots \\
\lambda(\bm{u}_N)
\end{bmatrix} = 
\begin{bmatrix}
1 & x_1 & y_1 & x_1^2 & x_1 y_1 & y_1^2 \\
1 & x_2 & y_2 & x_2^2 & x_2 y_2 & y_2^2 \\
\vdots & \vdots & \vdots & \vdots & \vdots & \vdots \\
1 & x_N & y_N & x_N^2 & x_N y_N & y_N^2 \\
\end{bmatrix},
\qquad
\mathcal{L} = 
\begin{bmatrix}
    L^{1}_{n} \\
    L^{2}_{n} \\
    \vdots \\
    L^{N}_{n}
\end{bmatrix}
=
\begin{bmatrix}
    L^{1}_{1} & L^{1}_{2} & \cdots & L^{1}_{M}  \\
    L^{2}_{1} & L^{2}_{2} & \cdots & L^{2}_{M} \\
    \vdots & \vdots & \ddots & \vdots \\
    L^{N}_{1} & L^{N}_{2} & \cdots & L^{N}_{M}
\end{bmatrix}
\end{equation}
The coefficient matrix $\Theta$ is obtained by fitting the known RK parameters for the six crystallographic directions (i.e., $i = 1, \dots, N, \enspace N = 6$) \jmr{given in \autoref{rk-coeffs}}. The $\Theta$ matrix can be used to obtain the unknown RK parameters ($\bm{L}_{n} = [L_1, L_2, L_3, L_4, L_5]$) for any new arbitrary direction $\kcap_{0}$ can be obtained by $\bm{L}_{n}(\kcap_{0}) = \lambda (\bm{u}_{0}) \Theta$.
\begin{table}[h]
\centering
\caption{$\mathbf{k}$ vectors and corresponding Redlich--Kister coefficients.}
\label{rk-coeffs}
\begin{tabular}{cccccccc}
\toprule
\multicolumn{3}{c}{\textbf{$\mathbf{k}$ vector}} & \multicolumn{5}{c}{\textbf{Redlich--Kister coefficients}} \\
\cmidrule(lr){1-3} \cmidrule(lr){4-8}
$k_x$ & $k_y$ & $k_z$ & $L_1$ & $L_2$ & $L_3$ & $L_4$ & $L_5$ \\
\midrule
0 & 0 & 1 & 0.0327 & 0.0258 & 0.0202 & 0.0225 & 0.0155 \\
1 & 1 & 1 & 0.0714 & 0.0407 & 0.0229 & 0.0152 & 0.0079 \\
0 & 1 & 1 & 0.0701 & 0.0329 & 0.0156 & 0.0087 & 0.0040 \\
2 & 0 & 1 & 0.0617 & 0.0319 & 0.0167 & 0.0105 & 0.0053 \\
2 & 2 & 1 & 0.0712 & 0.0376 & 0.0198 & 0.0129 & 0.0068 \\
3 & 1 & 1 & 0.0599 & 0.0341 & 0.0187 & 0.0115 & 0.0056 \\
\bottomrule
\end{tabular}
\end{table}

\pagebreak

\section{Details of the cluster expansion model}
\begin{figure}[h]
\centering
\includegraphics[width=\textwidth]{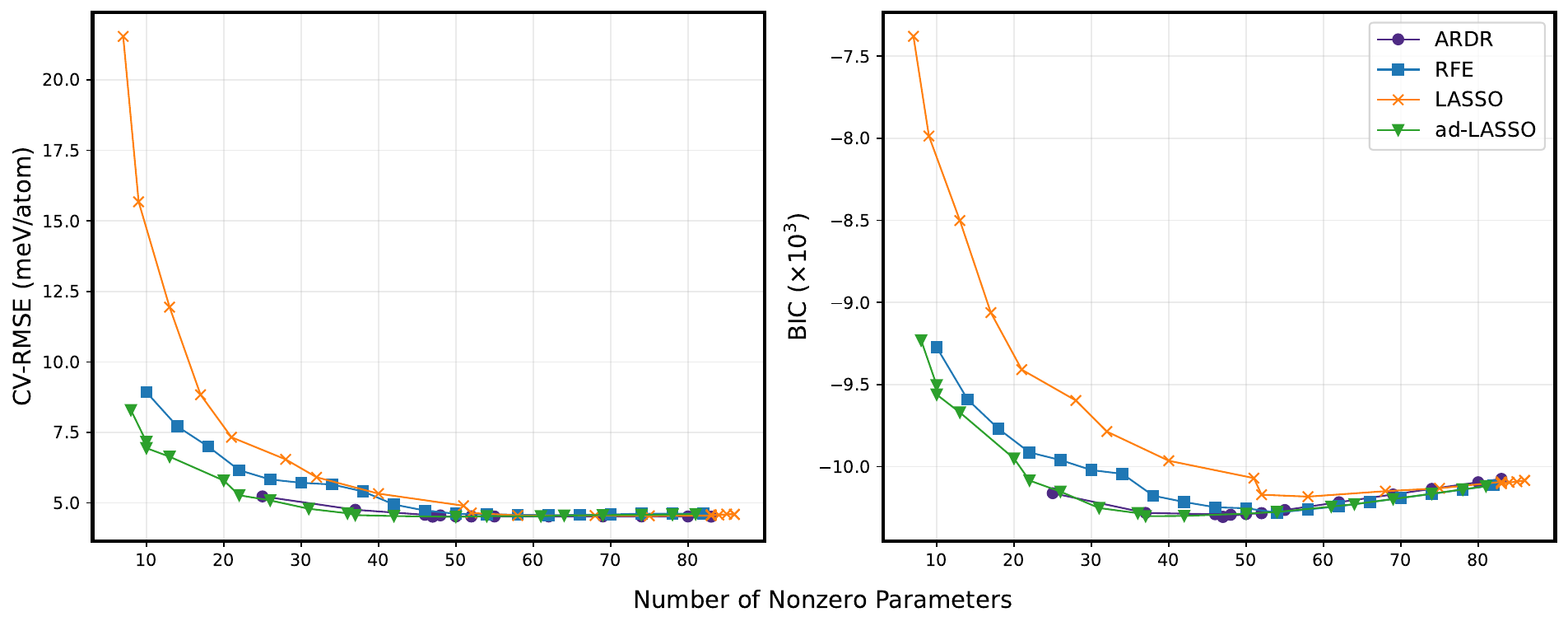}\vspace{-10pt}
\caption{Different fitting algorithms tested for training the CE model.}\label{fitting}
\end{figure}

\begin{figure}[h]
\centering
\includegraphics[width=\textwidth]{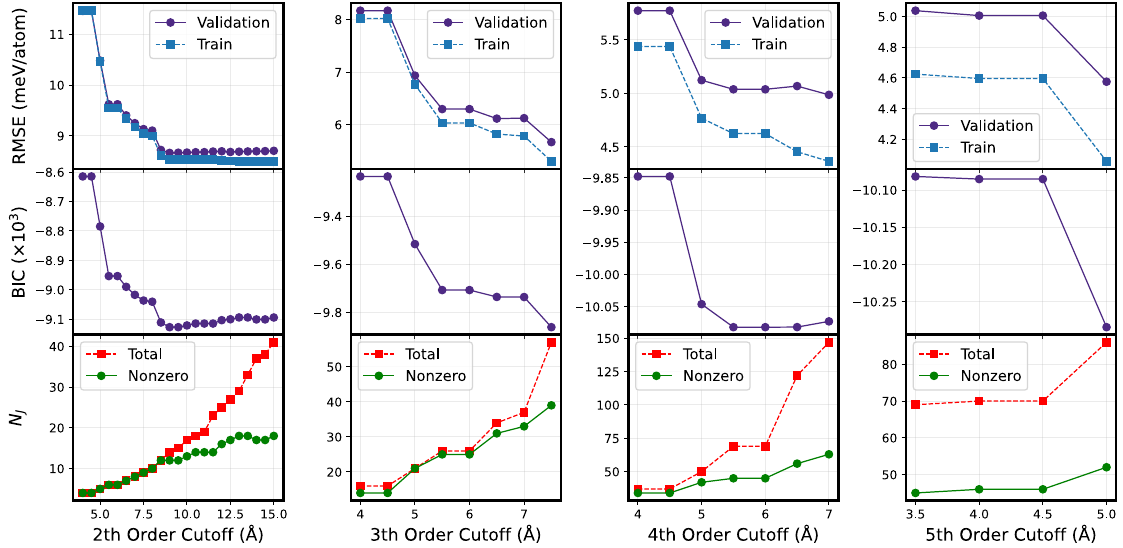}\vspace{-10pt}
\caption{Error metrics cross-validation RMSE and BIC, along with the total number of parameters ($N_J$) in the CE model, are plotted for clusters of different orders.}\label{cluster_cutoff} 
\end{figure}

\ar{The structures relaxed using DFT deviate from their ideal lattice positions post-relaxation, which is contradictory to the underlying assumption of the cluster expansion (CE) model that the atoms occupy fixed lattice positions. Therefore, we compute the displacement of atoms in the relaxed structures from the closest mapped position in the reference (primitive FCC) structure and \mjw{restrict our training set to only} structures that have a maximum displacement less than $0.4$ \AA\ and an average displacement less than $0.2$ \AA. This gives us a training dataset of 967 structures up to a supercell size of 10 atoms that meet the displacement criteria, with $631$ structures within the Rh composition range of $0.2\leq\comp\leq0.8$ and $336$ structures within the composition range of $0<\comp<0.2$ and $0.8<\comp<1.0$}. \ar{The training data is split into test data ($90\%$) and validation data ($10\%$) to perform model fitting. The CE model was obtained by fitting the DFT calculated mixing energy data to $\mathcal{H} - \mathcal{E_{CS}}$ using various fitting algorithms with arbitrary cutoffs. Initially, the CE model was fitted with arbitrarily chosen cluster cutoffs (close to and slightly greater than the final optimized cutoffs) to determine the best-fitting algorithm. The various fitting algorithms tested are: Automatic Relevance Determination Regression (ARDR), Recursive Feature Elimination (RFE), Least Absolute Shrinkage and Selection Operator (LASSO), and adaptive LASSO (ad-LASSO) \cite{fransson2020efficient}. The Cross-Validation Root Mean Square Error (RMSE) and Bayesian Information Criterion (BIC) metrics were used to determine both the fitting algorithm (\autoref{fitting}) and the optimal cluster cutoff (\autoref{cluster_cutoff}). We also track the total number of parameters in the model, denoted by $N_J$. The Automatic Relevance Determination Regression (ARDR) \cite{fransson2020efficient} method was found to be the optimal algorithm.} 

Thereafter, we selected the best-fitting algorithm (ARDR) to determine the optimized values of cluster cutoff radii. We obtained the optimal cluster cutoff values of $9.0$ \AA, $6.5$ \AA, $6.0$ \AA, and $5.0$ \AA\ for second, third, fourth, and fifth order clusters, respectively. With the optimized cutoffs and fitting algorithm, we trained the CE model. We obtained an optimized model with a Cross Validation Root Mean Square Error (CV-RMSE) of $4.1$ meV/atom for the training data and $4.7$ meV/atom for the validation data. \mjw{The optimized model has 12, 20, 35, and 17 non-zero ECIs corresponding to second, third, fourth, and fifth order clusters for a total of 84 parameters}

\begin{figure}[H]
\centering
\includegraphics[width=\textwidth]{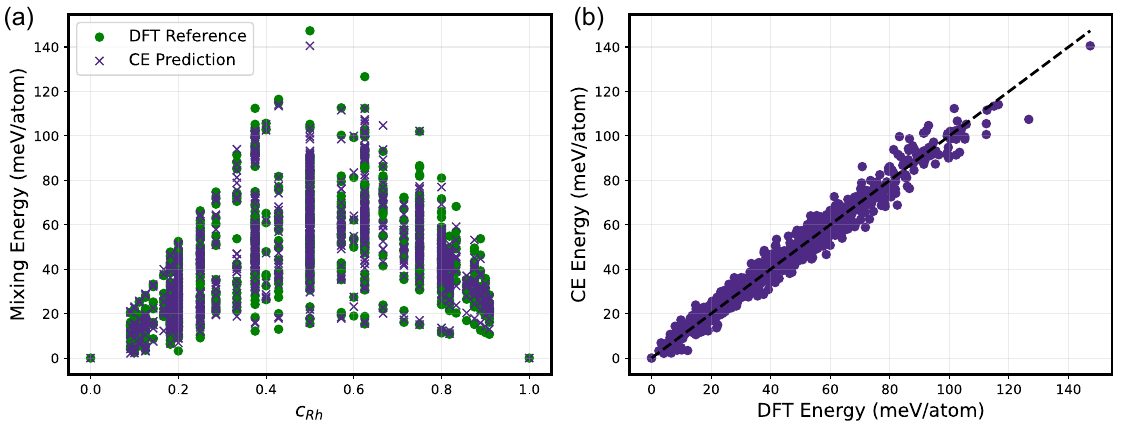}\vspace{-10pt}
\caption{Comparison of the CE prediction with reference (DFT) data. (a) The scatter plot of energies of all the training structures, along with the CE prediction. (b) The parity plot of DFT reference and CE prediction data.}\label{ce_prediction}
\end{figure}

We also compared the DFT-calculated mixing energies and the CE-predicted mixing energies to demonstrate the accurate predictions of the CE method, albeit in a computationally inexpensive manner (\autoref{ce_prediction}). \ar{Based on the \mjw{excellent} CV-RMSE values for both training and validation, less than $5$ meV/atom, along with the $R^2$ score of $0.96$, the model is within an acceptable level of accuracy for MC simulations.}

\begin{figure}[H]
\centering
\includegraphics[width=\textwidth]{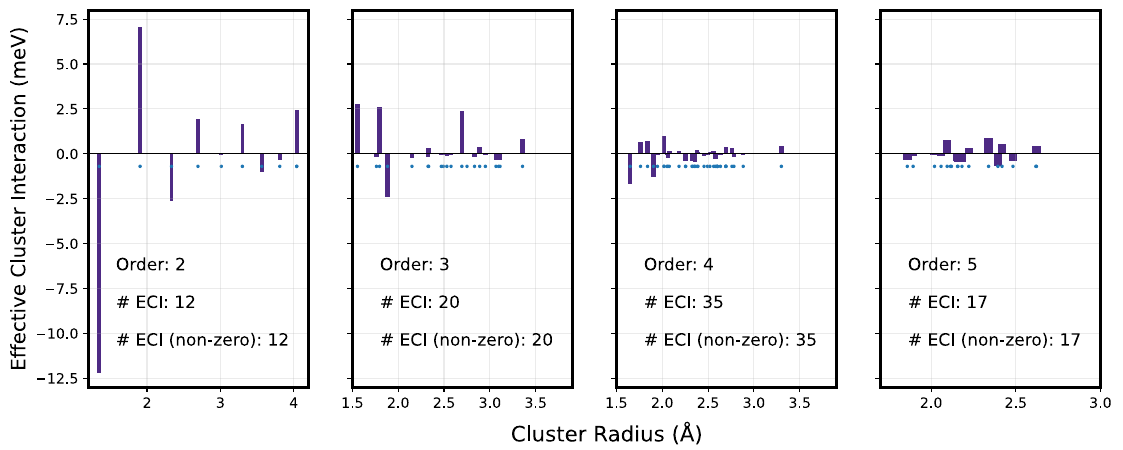}\vspace{-10pt}
\caption{Effective cluster interactions for clusters of various orders show a decaying trend with increasing cluster radius.}\label{ecis}
\end{figure}

\ar{The effective cluster interaction (ECI) parameters of the optimized CE model are shown in \autoref{ecis}, following a common trend where the magnitude of the parameters decay with increasing cluster radius. The zerolet term ($J_0$) and singlet ($1$-body cluster) being $59.5$ meV and $23.5$ meV are obtained by constraining the model to produce the mixing energies of pure phases exactly.} The relative magnitude of ECIs also influences the contribution of clusters of different orders toward the CE model.
\pagebreak

\section{Details of Monte Carlo simulations}
\begin{figure}[h]
	\centering
	\includegraphics[width=\textwidth]{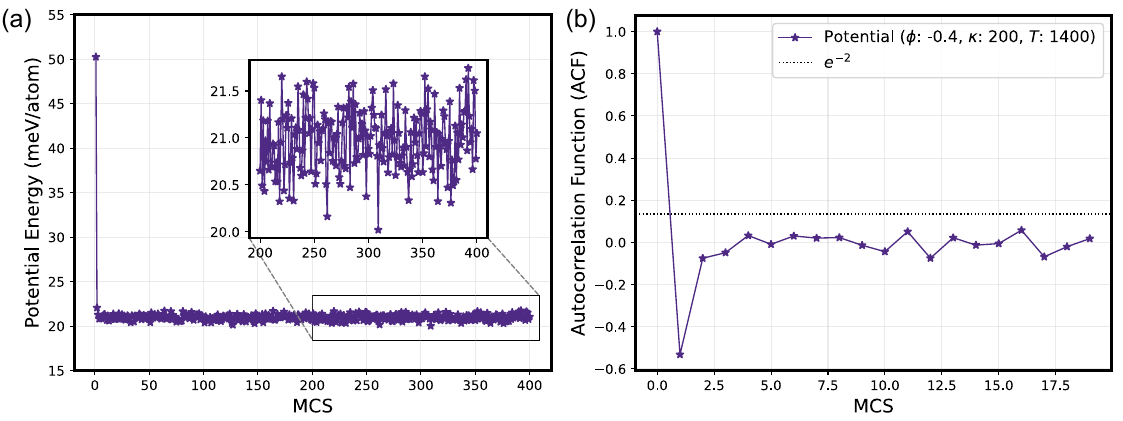}\vspace{-10pt}
	\caption{\ar{(a) The potential energy (mixing energy in this case) calculated from MC simulations for $\phi=-0.4$, $\kappa=200$, and $t=1400$K. (b) The autocorrelation function of the potential.}}\label{pot_energy}
\end{figure}

The potential energy for the Monte Carlo (MC) simulation (in this case, the mixing energy) is shown in \autoref{pot_energy}(a). From the inset in \autoref{pot_energy}(a), it can be seen that the potential energy variation is in the order of 2 meV/atom per MCS post equilibration. While sampling the configuration space, we want to ensure that the samples are sufficiently far apart so that they are uncorrelated. For this, we calculated the autocorrelation function (ACF) for potential energy as shown in \autoref{pot_energy}(b) and calculated the correlation length as the minimum number of steps after which the ACF value goes below $e^{-2}$.

\begin{figure}[h]
    \centering
    \includegraphics[width=\textwidth]{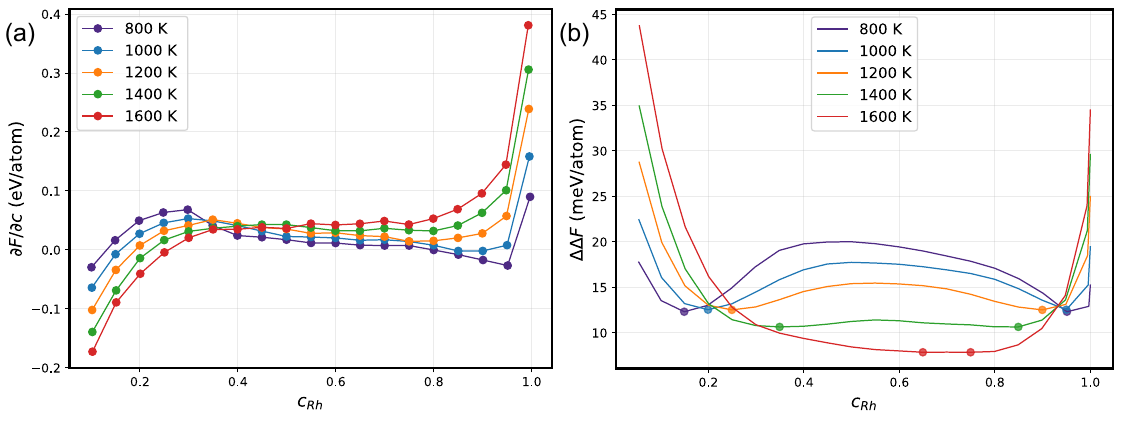}\vspace{-10pt}
    \caption{(a) Free energy derivative obtained from VCSGC ensemble MC simulations. (b) Relative free energy of mixing.}\label{free_energy}
\end{figure}

\begin{figure}[h]
	\centering
	\includegraphics[width=0.5\textwidth]{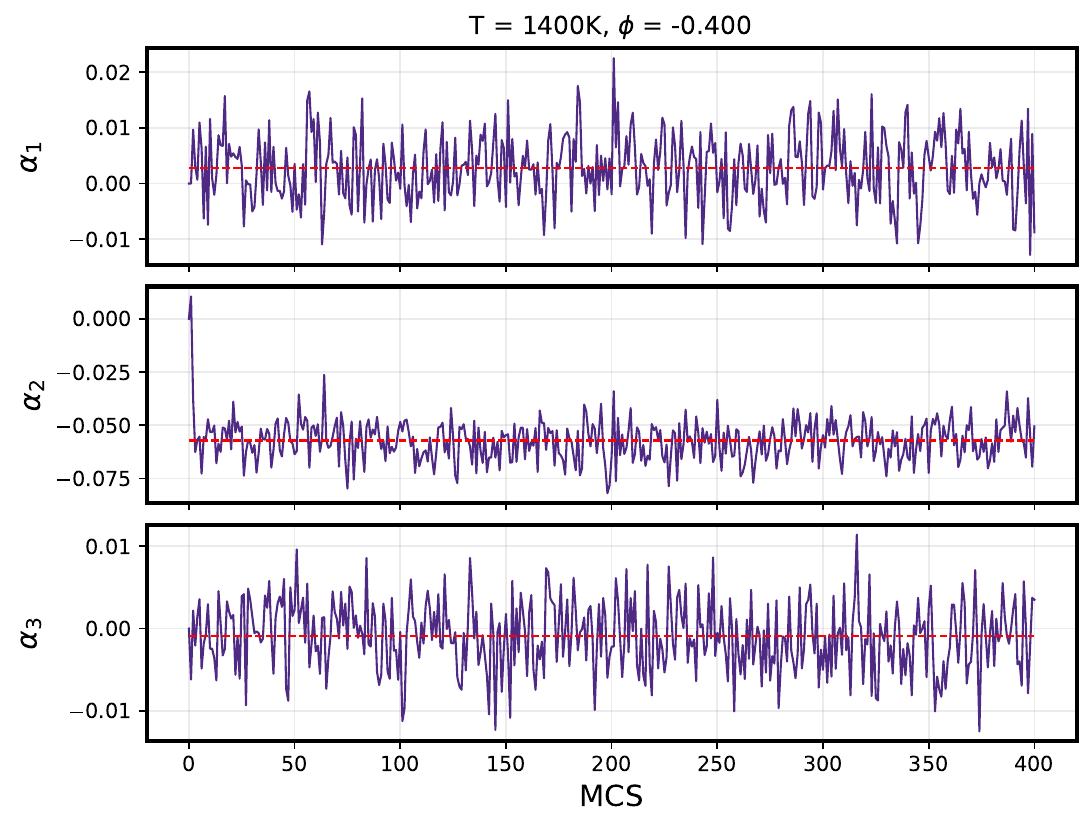}\vspace{-10pt}
	\caption{First three nearest neighbor Warren-Cowley SRO parameter distribution plotted against Monte Carlo Sweeps.}\label{sro_dist}
\end{figure}

The free energy derivative obtained from the MC simulation is shown in \autoref{free_energy}(a). Using the common tangent construction for the free energy plots derived via thermodynamic integration, we obtained the points corresponding to the incoherent phase boundary. These points are marked in the relative free energy of mixing plot as shown in \autoref{free_energy}(b).
We also obtained the Warren-Cowley chemical short-range order (SRO) parameters calculated at each Monte Carlo step. We plotted them for the entire simulation length of 400 Monte Carlo Sweeps (MCS), as shown in \autoref{sro_dist}.

The central red line in \autoref{sro_dist} represents the average value calculated over the entire simulation range. However, while reporting the ensemble average, we calculated the average over the last 200 MCS (discarding the initial 200 MCS equilibration period). To ensure sufficient sampling of the configuration space such that the ensemble average of the SRO parameter approximates the true population mean value, we computed the standard error of the mean ($\epsilon$) and the relative standard error.
\begin{figure}[H]
	\centering
	\includegraphics[width=\textwidth]{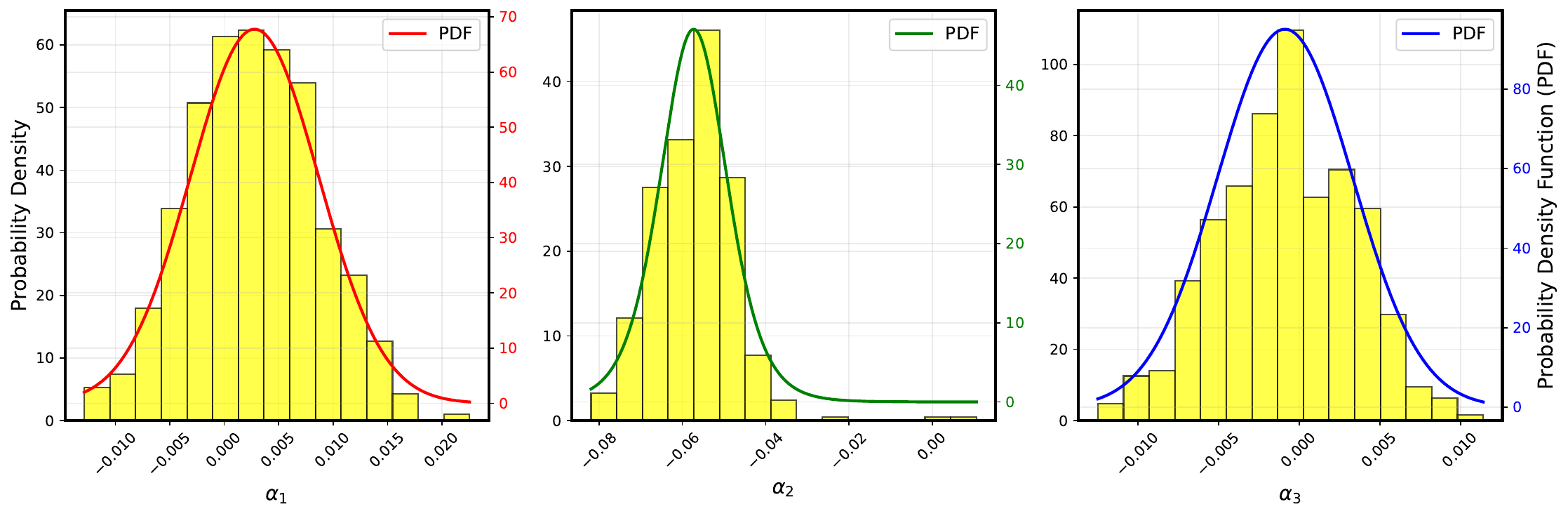}\vspace{-10pt}
	\caption{SRO parameter distribution fitted to Student's t-distribution.}\label{sro_distribution}
\end{figure}
The standard error of the mean was calculated by fitting a Student's t-distribution to the SRO parameter histogram (shown in \autoref{sro_distribution} for $\phi=-0.4$, $\kappa=200$, and $T=1400K$) with the factor $t*$ corresponding to the $95\%$ confidence interval. Thereafter, the standard error of mean can be computed as 
$\epsilon = {\sigma t*}/ \sqrt{\mathcal{N}}$, 
where $\sigma$ is the standard deviation of the SRO distribution and $\mathcal{N}$ is the total number of uncorrelated samples. 

\begin{figure}[H]
	\centering
	\includegraphics[width=\textwidth]{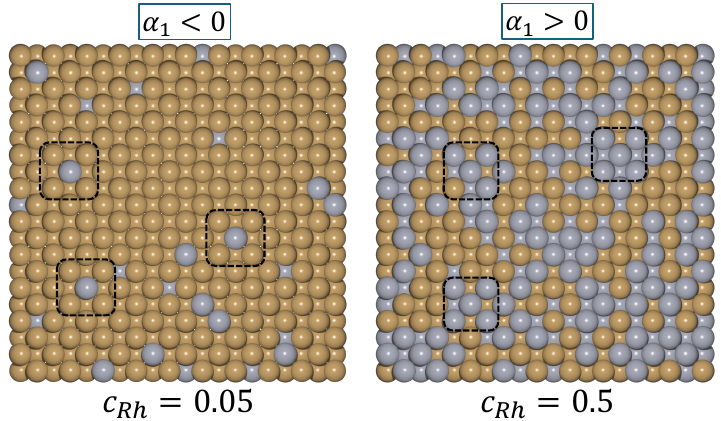}\vspace{-10pt}
	\caption{Snapshots of MC simulation supercell for two different compositions.}\label{supercell_snapshots}
\end{figure}

The relative standard error is $\delta = {\epsilon}/{\mu}$, where $\mu$ is the calculated ensemble average of the distribution (not the true population mean).
\rev{We have also provided atomic snapshots of the Monte Carlo simulation cell for two different Rh compositions to visualize SRO and understand how the parameters reflect the atomic configurations at a given temperature and composition. From \autoref{supercell_snapshots} it can be seen that for $\comp = 0.05$ the 1st nearest neighbor (NN) SRO parameter $\alpha_1$ is negative, which represents pockets of atomic configuration where the Rh atom (represented by gray atoms) is surrounded by Cu atoms. Similarly, for $\comp=0.5$, the SRO parameter is positive, which indicates the presence of like atoms in the first coordination shell.}
\pagebreak

\section{Percolation threshold variation due to chemical short-range order}
\begin{figure}[H]
	\centering
	\includegraphics[width=0.5\textwidth]{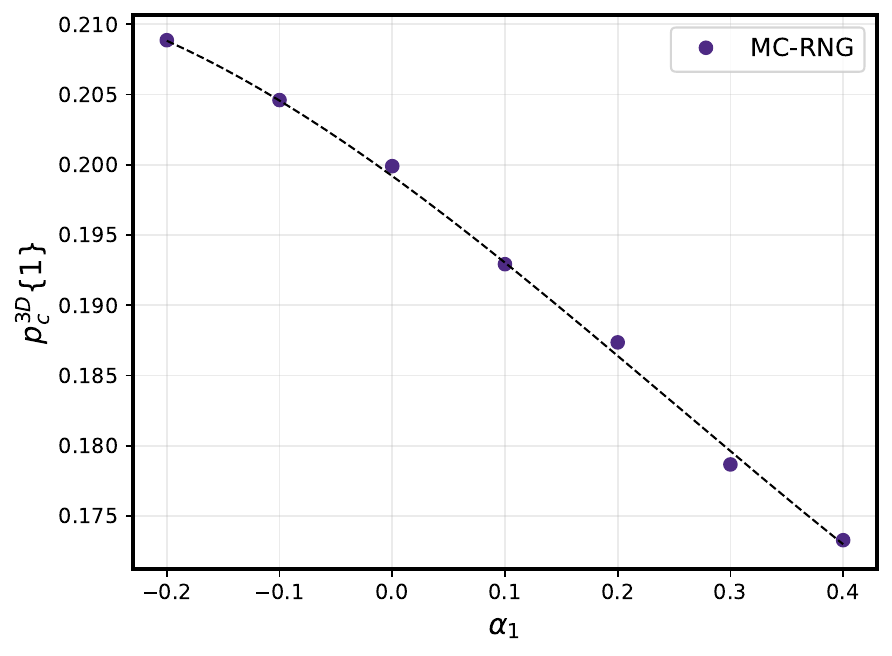}\vspace{-10pt}
	\caption{Percolation threshold variation for different SRO parameters. Reproduced from Ref.\ \cite{roy2024effect}.}\label{pc_fit}
\end{figure}

We obtained the 1st nearest neighbor (NN) percolation threshold diagrams from the calculated SRO parameters using the polynomial fit to the data in \autoref{pc_fit} reproduced from Ref.\ \cite{roy2024effect}. The data points are obtained using the Large Cell Monte Carlo Renormalization Group method as outlined in Ref. \cite{roy2024effect}.
\pagebreak
\bibliographystyle{elsarticle-num} 
\bibliography{refs_paper2.bib}